\documentclass[letter]{aa}  
\usepackage{epstopdf}
\usepackage{graphicx}
\usepackage{color,soul}
\usepackage{txfonts}
\begin{document}

   \title{The Extreme Ultraviolet and X-Ray Sun in Time:\\ High-Energy Evolutionary Tracks
          of a Solar-Like Star}


   \titlerunning{The X-ray and EUV Sun in Time}
   \author{Lin Tu
          \inst{1}
          \and
          Colin P. Johnstone
	  \inst{1}
	  \and
	  Manuel G\"udel
	  \inst{1}
	  \and
	 Helmut Lammer
	 \inst{2}
	  }

   \institute{$^1$Department of Astrophysics, University of Vienna,
       T\"{u}rkenschanzstrasse 17, A-1180 Vienna\\
              \email{lin.tu@univie.ac.at, colin.johnstone@univie.ac.at, manuel.guedel@univie.ac.at}\\
              $^2$              Space Research Institute, Austrian Academy of Sciences, Schmiedlstrasse 6, A-8042 Graz, Austria\\
              \email{helmut.lammer@oeaw.ac.at}             }

   \date{Received August 10, 2014; accepted ...}

 
  \abstract
   {}
   {We aim to describe the pre-main sequence  and main-sequence evolution of X-ray and extreme-ultaviolet 
   radiation of a solar mass star based on its rotational evolution starting with a realistic range of initial rotation rates.}
   {We derive evolutionary tracks of X-ray radiation based on  a rotational evolution model for solar mass stars and the rotation-activity relation. We compare these tracks to X-ray luminosity distributions of stars in clusters with different ages.}
   {We find agreement between the evolutionary tracks derived from rotation and the X-ray luminosity distributions from
   observations. Depending on the initial rotation rate, a star might remain at the X-ray saturation level for
   very different time periods, from $\approx 10$~Myr to $\approx$300~Myr for slow and fast rotators, respectively. }
   {Rotational evolution with a spread of initial conditions leads to a particularly wide distribution 
   of possible X-ray luminosities in the age range of 20--500~Myrs, before rotational convergence and therefore
   X-ray luminosity convergence sets in. This age range is crucial for the evolution of young planetary atmospheres
   and may thus lead to very different planetary evolution histories. }


   \maketitle
%
\section{Introduction}
High-energy radiation from solar-like main-sequence (MS) stars decays in time
owing to stellar spin-down. 
The early Sun's X-ray  ($\approx 1-100$~\AA) and extreme-ultraviolet ($\approx 100-900$~\AA) emissions
could thus have exceeded the present-day Sun's level by orders of magnitude \citep{ribas05}. 
By driving atmospheric erosion, such extreme radiation levels were critically important for both the primordial hydrogen atmospheres (e.g., \citealt{lammer14}) and
the secondary nitrogen atmospheres \citep{lichtenegger10} of solar system planets. As a consequence of higher solar 
activity levels, stronger winds would have added to atmospheric mass loss through  
non-thermal processes such as ion pick-up (\citealt{kislyakova13}; \citealt{2014A&A...562A.116K}).

Magnetic activity is strongly coupled to rotation via a stellar dynamo, such that the total X-ray 
luminosity decays with increasing rotation period, $P$, as $L_{\rm X} \propto P^{-3}$ to 
$\propto P^{-2}$; for small $P$, $L_{\rm X}$ saturates at  $L_{\rm X} \approx 10^{-3}L_{\rm bol}$ 
($L_{\rm bol}$ being the stellar bolometric luminosity; see \citealt{wright11}=W11). Since older (>1~Gyr) stars
spin down in time approximately as $P \propto t^{0.5}$ \citep{skumanich72}, $L_{\rm X}$ decays  
as $L_{\rm X} \propto t^{-1.5}$ \citep{guedel97}. This evolutionary trend has commonly been 
formulated using regression fits to $L_{\rm X}$ of stars with known ages, typically 
starting at the saturation level close to the zero-age main-sequence 
(ZAMS; see, e.g.,  \citealt{guedel97, ribas05}).

However, stars in young stellar clusters have a wide spread in rotation rates, $\Omega$, in particular 
at ages younger than 500~Myr before they converge to a unique mass-dependent
value \citep{soderblom93}. As a consequence, $L_{\rm X}$ values also scatter over a wide range
among such stars (e.g., \citealt{stauffer94}), and the age at which stars fall out of saturation 
 depends on the star's initial $\Omega$. Given the importance of 
high-energy radiation in this age range for planetary atmosphere evolution, a unique
description with a single radiation decay law is problematic and needs
to be replaced by a description of the $L_\textbf{X}$ distribution and its long-term
evolution (\citealt{2008A&A...477..309P}; \citealt{2015arXiv150307494J}), spanning a wide range of possible evolutionary tracks for stars with different initial $\Omega$.

In this Letter, we use a rotational evolution model to predict such 
luminosity distributions as a function of age, based on a range of initial $\Omega$, and 
we show that these predictions agree with the observed time-dependent scatter of $L_{\rm X}$.
We derive a radiative evolution model 
based on the  full range of rotation histories for a solar-mass star, and thus
find a description of the possible past histories of our own Sun, which is useful to model the corresponding
evolution of solar-system planetary atmospheres. 
This Letter is an extension of \citet{2015arXiv150307494J}, who similarly estimated evolutionary tracks for wind properties.
In this Letter, we concentrate mostly on 1~M$_\odot$ stars, and will
extend this to other stellar masses in future work.

\begin{figure*}
\centering
\includegraphics[angle=0,width=0.49\textwidth]{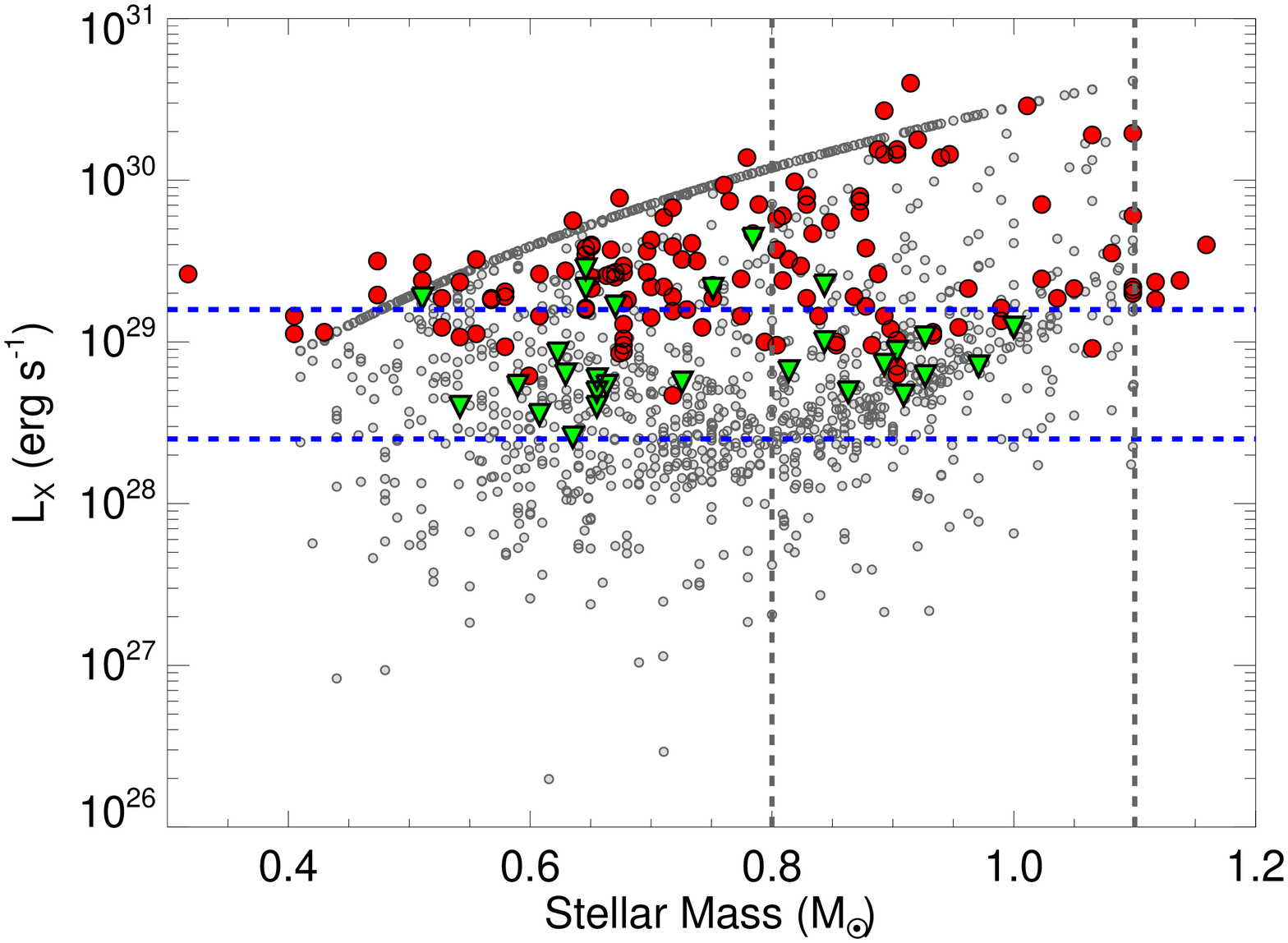}
\includegraphics[angle=0,width=0.49\textwidth]{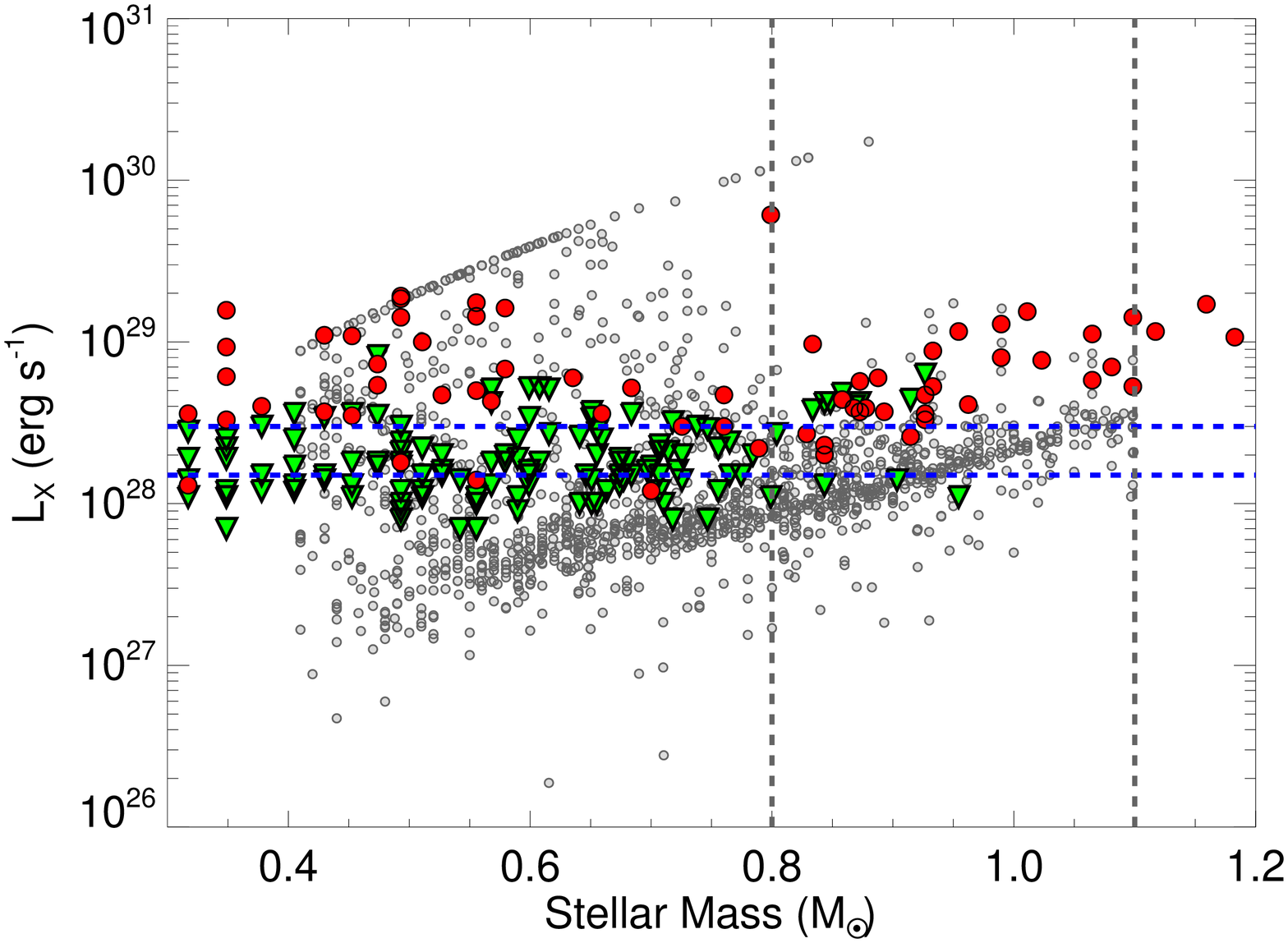}
\caption{ 
Comparisons between observed and predicted distributions of $L_\text{X}$ at ages of 150~Myr (\emph{left}) and 620~Myr (\emph{right}). 
The grey symbols show predicted distributions calculated using Eqn.~\ref{eqn:wrightlaw} and the distributions of rotation rates derived by \citet{2015arXiv150307494J}.
The red circles and green triangles show detections and upper limits for stars in the Pleiades (\emph{left}) and Hyades (\emph{right}). 
The horizontal blue lines show detection thresholds.
The upper line of stars in the theoretical distributions is caused by
the stars whose rotation rates lie above the mass dependent saturation threshold.
}
\label{fig1}
\end{figure*}

\section{Rotation and Radiation Models} \label{sect:model}

As in previous studies (e.g. \citealt{gallet13}), we constrain our rotation models by assuming that the percentiles of the rotational distributions for star clusters with different ages can be combined to estimate the time evolution of a star's rotation rate.
We consider only stars in the mass range
0.9~M$_\odot$ to 1.1~M$_\odot$.
\citet{2015arXiv150307494J} collected measured rotation periods of over 2000 stars in clusters of ages 150, 550, and 1000 Myr on
the MS, giving observational constraints on the percentiles at
these ages (with 230, 134, and 36 stars, respectively at the considered ages).
We use additional constraints for pre-main sequence (PMS) rotation from the
$\approx$2~Myr cluster NGC~6530 (28~stars; \citealt{2012ApJ...747...51H})
and the $\approx$~12~Myr cluster h~Per (117 stars;
\citealt{2013A&A...560A..13M}). 
For NGC~6530, the 10th, 50th, 
and 90th percentiles are at 2.7$\Omega_\odot$, 6.2$\Omega_\odot$, and 35.1$\Omega_\odot$, respectively (assuming $\Omega_\odot = 2.9 \times 10^{-6}$~rad~s$^{-1}$), 
and for h~Per, they are at 3.4$\Omega_\odot$, 8.4$\Omega_\odot$, and 76.0$\Omega_\odot$, respectively.

We use an extension of the rotational evolution model of \citet{2015arXiv150307494J}. 
For the wind torque, we use the formula derived by \citet{2012ApJ...754L..26M} which relates the wind torque to stellar parameters, the star's dipole field strength, $B_\text{dip}$, and the wind mass loss rate, $\dot{M}_\star$.
We assume that both $B_\text{dip}$ and $\dot{M}_\star$ saturate at a Rossby number of \mbox{$Ro=0.13$}, as suggested by the saturation of X-ray emission (W11), where $Ro = P_\text{rot} / \tau_\star$ and $\tau_\star$ is the convective turnover time.
For $\dot{M}_\star$, we use the scaling law derived by Johnstone et al. (2015), which is derived by fitting the rotational evolution model to observational constraints.
We modify the scaling law by relating $\dot{M}_\star$ to $Ro$; this allows us to take into account the change in $\tau_\star$ on the PMS. 
Since we only consider solar mass stars, the scaling relation derived by Johnstone et al. (2015) can be rewritten as $\dot{M}_\star \propto R_\star^2 Ro^{-a}$.
We find \mbox{$a=2$} provides a good fit to the observational constraints (which is larger than the value of \mbox{$a=1.33$} found by Johnstone et al.~2015).  
For $B_\text{dip}$, we use the scaling law derived by \citet{2014MNRAS.441.2361V} of $B_{\text{dip}} \propto Ro^{-1.32}$.

To reproduce the spin up on the PMS due to the decrease in the stellar moment of inertia, previous studies have found that core-envelope decoupling must be included (\citealt{1997ApJ...480..303K}).
We use the core-envelope decoupling model described by \citet{2015arXiv150205801G} and adopt coupling timescales of 30~Myr, 20~Myr, and 10~Myr for the 10th, 50th and 90th percentile tracks, respectively, which we find give us good agreement between the rotational evolution
model and the observations.
Finally, we assume that during the first few million years, the stellar rotation rates do not evolve with time due to `disk-locking', i.e. magnetic interactions with the circumstellar disk. 
We assume disk-locking timescales of 10~Myr, 5~Myr, and 2~Myr for the 10th, 50th and 90th percentile tracks, respectively. 

To predict $L_{\rm X}$ along the rotation tracks, we use the relation derived from MS 
stars by W11, 
\begin{equation} \label{eqn:wrightlaw}
R_\text{X} = \left \{
\begin{array}{ll}
C Ro^\beta, & \text{if } Ro \ge Ro_{\text{sat}},\\
R_{\text{X},\text{sat}}, & \text{if } Ro \le Ro_{\text{sat}},\\
\end{array} \right.
\end{equation}
where \mbox{$Ro_\text{sat} = 0.13$} is the saturation Rossby number, \mbox{$R_\text{X} = L_\text{X} / L_\text{bol}$}, and \mbox{$R_{\text{X}, \text{sat}} = 10^{-3.13}$} is the saturation $R_\text{X}$ value. 
We use $\beta=-2.7$ (W11).
We assume that this relation can be used on the PMS if the evolution of $L_{\rm bol}$ and $\tau_\star$ are treated correctly. 
\citet{2011A&A...532A...6S} derived a power law to convert $L_\text{X}$ (5-100\AA) into EUV luminosity, $L_\text{EUV}$ (100-920\AA), of $\log L_\text{EUV} = 4.8 + 0.86 \log L_\text{X}$, where $L_\text{X}$ and $L_\text{EUV}$ are in erg~s$^{-1}$.

To calculate the evolution of the stellar radius, $L_{\rm bol}$, the moment of inertia, and $\tau_\star$, we use the stellar evolution models of \citet{2013ApJ...776...87S}.
However, their $\tau_\star$ values are approximately a factor of two above those of W11 for 1 M$_\odot$ stars; we therefore normalize $\tau_\star$ at all ages such that the MS value is consistent with Eqn.~\ref{eqn:wrightlaw}.

\begin{figure*}
\centering
\includegraphics[trim=12mm 5mm 12mm 10mm, clip=true, angle=0,width=0.49\textwidth]{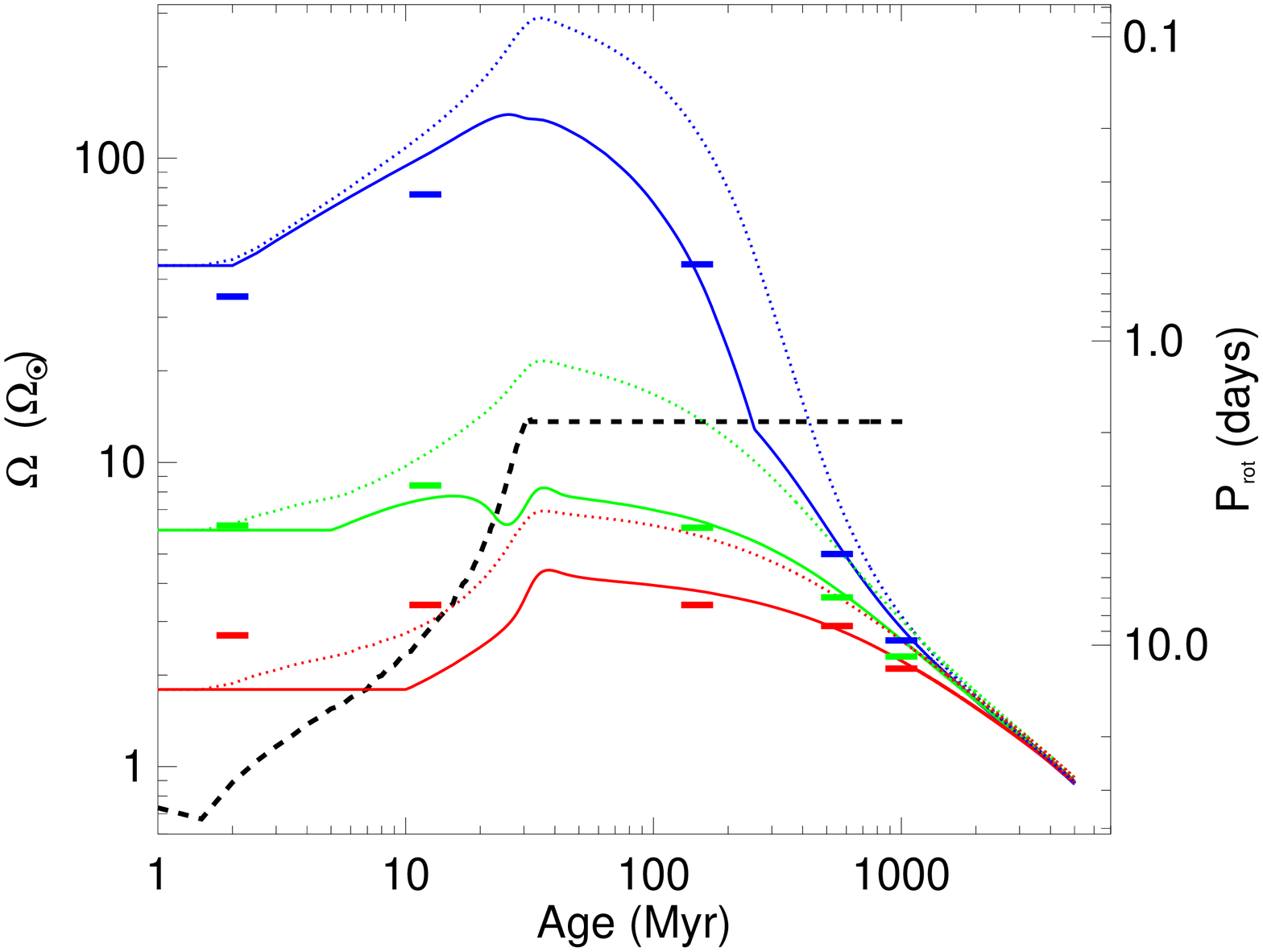}
\includegraphics[trim=12mm 5mm 12mm 10mm, clip=true, angle=0,width=0.49\textwidth]{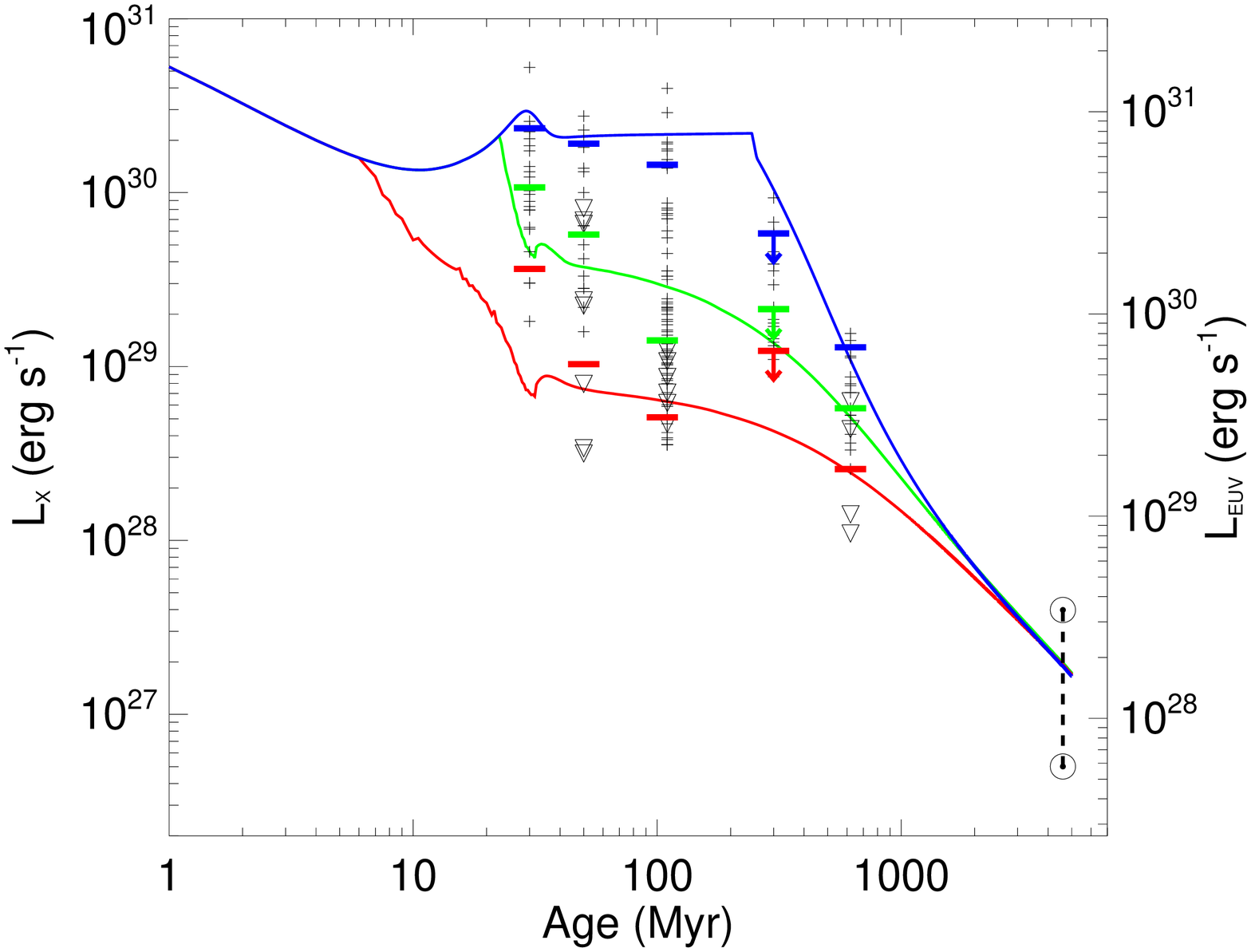}
\caption{ 
{\it Left (a):} Predicted rotational evolution tracks for stars at the 10th (\emph{red}), 50th (\emph{green}), and 90th (\emph{blue}) percentiles of the rotational distribution.
The solid and dotted lines show the envelope and core rotational evolution, respectively, and the horizontal solid lines show the observational constraints on the percentiles. 
The dashed black line shows the time dependent saturation threshold for $\dot{M}$, $B_\text{dip}$, and $L_\text{X}$, calculated assuming a constant saturation $Ro$ and the $\tau_\star$ values of \citet{2013ApJ...776...87S}. 
{\it Right (b):} Predicted $L_{\rm X}$ along each of our rotation tracks and comparisons to observed LX values of single stars in
several clusters, with upper limits shown by $\bigtriangledown$ symbols. The solid horizontal lines show the 10th, 50th, and 90th percentiles of the observed distributions of $L_\text{X}$ at each age, calculated by counting upper limits as detections. The two solar symbols at 4.5~Gyr show the range of $L_\text{X}$ for the Sun over the course of the solar cycle. The scale on the right y-axis shows the associated $L_\text{EUV}$.  
}\label{fig:tracks}
\end{figure*}

\section{X-ray Observations} \label{sect:obs}

To test our predictions for X-ray distributions at each age, we collect $L_\text{X}$  values of single stars from ROSAT, XMM-Newton, and Chandra of open clusters with ages from 30 Myr to 620 Myr. 
The clusters 
are NGC~2547 (30~Myr; \citealt{Jeffries2006}), $\alpha$ Persei (50~Myr; \citealt{Prosser1998}), NGC~2451 (50~Myr; \citealt{Huensch2003}), Blanco~I (50~Myr; \citealt{2003A&A...399..919P}), Pleiades (100~Myr; \citealt{Micela1999a}; \citealt{stauffer94}), NGC~2516 (110~Myr; \citealt{Pillitteri2006}; \citealt{Jeffries1997}), NGC~6475 (300~Myr; \citealt{1995AJ....110.1229P}), and Hyades (620~Myr; \citealt{Stern1995}). 
For NGC~6475, since no optical catalogue was available, \citet{1995AJ....110.1229P} did not report upper limits for the non-detected stars and therefore the percentiles for the distribution of $L_\text{X}$ should be considered upper limits. 
For all MS clusters, except Blanco~I where masses were given, we derive
masses by converting from $(B-V)_0$ using a relation derived from the \citet{2007ApJ...671.1640A} stellar evolution models.
For the PMS cluster NGC 2547, we calculated masses using the \citet{2000A&A...358..593S} models.
Since we use these X-ray observations only to compare to our
predictions from rotation, we do not attempt to homogenise the
$M_\star$ and $L_\text{X}$ determinations for each cluster.
Our quantitative determinations of the $L_\text{X}$ tracks are based on
the relation from W11 where such homogenisation was done.

%
%
%
%
%
%
%
%
%
%
%

\section{Results}\label{sect:res}

\citet{2015arXiv150307494J} combined rotation period measurements of four young clusters with ages of $\sim$150~Myr and used a rotational evolution model to predict the evolution of the resulting distribution of $\Omega$ on the MS. 
The sample contains 1556 stars in the 0.4-1.1~M$_\odot$ mass range.
In Fig.~\ref{fig1}, we show predictions for the distributions of $L_\text{X}$ based on these $\Omega$ distributions  at ages of 150~Myr and 620~Myr comparing with observed values in the Pleiades and Hyades, respectively.
There is good overall agreement, although intrinsic X-ray variability (typically factors of 2--3) introduces some additional scatter such as is visible for stars exceeding the
saturation threshold.

To predict the range of possible $L_{\rm X}$ evolution tracks, we calculate rotation models for solar mass stars at the 10th, 50th, and 90th percentiles of the $\Omega$ distributions, shown in Fig.~\ref{fig:tracks}a.
Our models fit well the observational constraints on the percentiles, except for a slight underestimation of the 10th percentile track in the first 20~Myr. 
This might cause us to underestimate the age when stars on the 10th percentile track come out of saturation by a few Myr.  
Fig.~\ref{fig:tracks}b shows predicted tracks for $L_\text{X}$ and $L_\text{EUV}$
together with observed $L_{\rm X}$ for stars in the $0.9-1.1~M_{\odot}$ range for each cluster listed in Sect.~\ref{sect:obs}.
Due to the low number of observations in NGC~2547 (30~Myr), we extend the mass range to $0.8-1.2~M_{\odot}$.
The tracks correspond very well to the observed percentiles in the individual clusters given the somewhat limited observational
samples. 
The solar $L_{\rm X}$ (6$\times 10 ^{26} - 5\times10 ^{27}$~erg~s$^{-1}$, \citet{1997JGR...102.1641A};  \citet{Peres2000};  \citet{Judge2003};)
has been included as well and fits our models excellently.

Stars on our rotation tracks drop out of saturation at $\approx$6~Myr (10th percentile, red), $\approx$20~Myr (50th, green),
and $\approx$300~Myr (90th, blue), i.e., either as young PMS stars, as near-ZAMS stars,
or as slightly evolved MS stars. The spread in $L_{\rm X}$ amounts to as much as 1.5 orders of magnitude for several 100 Myr.



Fig.~3 gives the age when a star falls out of saturation, $t_{\rm sat}$, as a function of initial 
$\Omega$, derived from our rotation model.
This ``saturation time'' can be approximated by 
\begin{equation}\label{tsat}
t_{\rm sat} = 2.9\Omega_0^{1.14},
\end{equation}
where $t_{\rm sat}$ is in Myr and $\Omega_0$ is the rotation rate at 1~Myr in units of the solar rotation rate. 
Assuming that the saturation level, $L_{\rm X, sat}\approx10^{-3.13} L_{\text{bol},\odot}$, is constant in time, which is approximately true, we obtain
\mbox{$\log L_{\rm X, sat} = 30.46$}. If we approximate $L_{\rm X}$ by a power law after $t_{\rm sat}$ (see Fig. 2b), we obtain, for a given $\Omega_0$,
\begin{equation}
L_{\rm X} = \left\{ 
  \begin{array}{l l}
     L_{\rm X, sat}, & \quad \text{if\ $t\le t_{\rm sat}$},\\[0.2cm]
     a t^b,          & \quad \text{if\ $t \ge  t_{\rm sat}$}.\\
  \end{array}
  \right.
\end{equation}
We require that the power law also fits the Sun with \mbox{$L_{\rm X, \odot} = 10^{27.2}$}~erg~s$^{-1}$ at an
age of $t_{\odot} = 4570$~Myr. We thus find
\begin{equation}
b^{-1}        =  0.35\log \Omega_0 - 0.98 , \hspace{4mm} 
a             = L_{\mathrm{X}, \odot}t_\mathrm{\odot}^{-b}.
\end{equation}
For the 10th, 50th, and 90th percentiles in $\Omega_0$, corresponding to $\Omega_0 \approx 1.8\Omega_{\odot}, 6.2\Omega_{\odot}$, 
and 45.6~$\Omega_{\odot}$ with $t_{\rm sat} \approx$~5.7~Myr, 23~Myr, and 226~Myr, respectively, we find 
\begin{equation}
\hbox{$
L_{\rm X} = \left\{ 
  \begin{array}{l}
     2.0\times 10^{31} t^{-1.12} \\[0.15cm]
     2.6\times 10^{32} t^{-1.42} \\[0.15cm]
     2.3\times 10^{36} t^{-2.50} \\[0.15cm]
  \end{array}
  \right.
L_{\rm EUV} = \left\{ 
  \begin{array}{l l}
     7.4\times 10^{31} t^{-0.96} & \quad \text{10th}\\[0.15cm]
     4.8\times 10^{32} t^{-1.22} & \quad \text{50th}\\[0.15cm]
     1.2\times 10^{36} t^{-2.15} & \quad \text{90th}\\[0.15cm]
  \end{array}
  \right.
$}\nonumber
\end{equation}
where luminosities are in erg~s$^{-1}$.
The slope of the median $L_{\rm X}$ track, $b = -1.42$, is very close to values reported from linear regression to the Sun in Time sample
\citep{guedel97, ribas05}.  These power-law fits, valid for $t > t_{\rm sat}$, thus describe the range
of possible evolutionary tracks for $L_{\rm X}$ and $L_{\rm EUV}$.
 
\begin{figure}
\includegraphics[angle=0,width=0.49\textwidth]{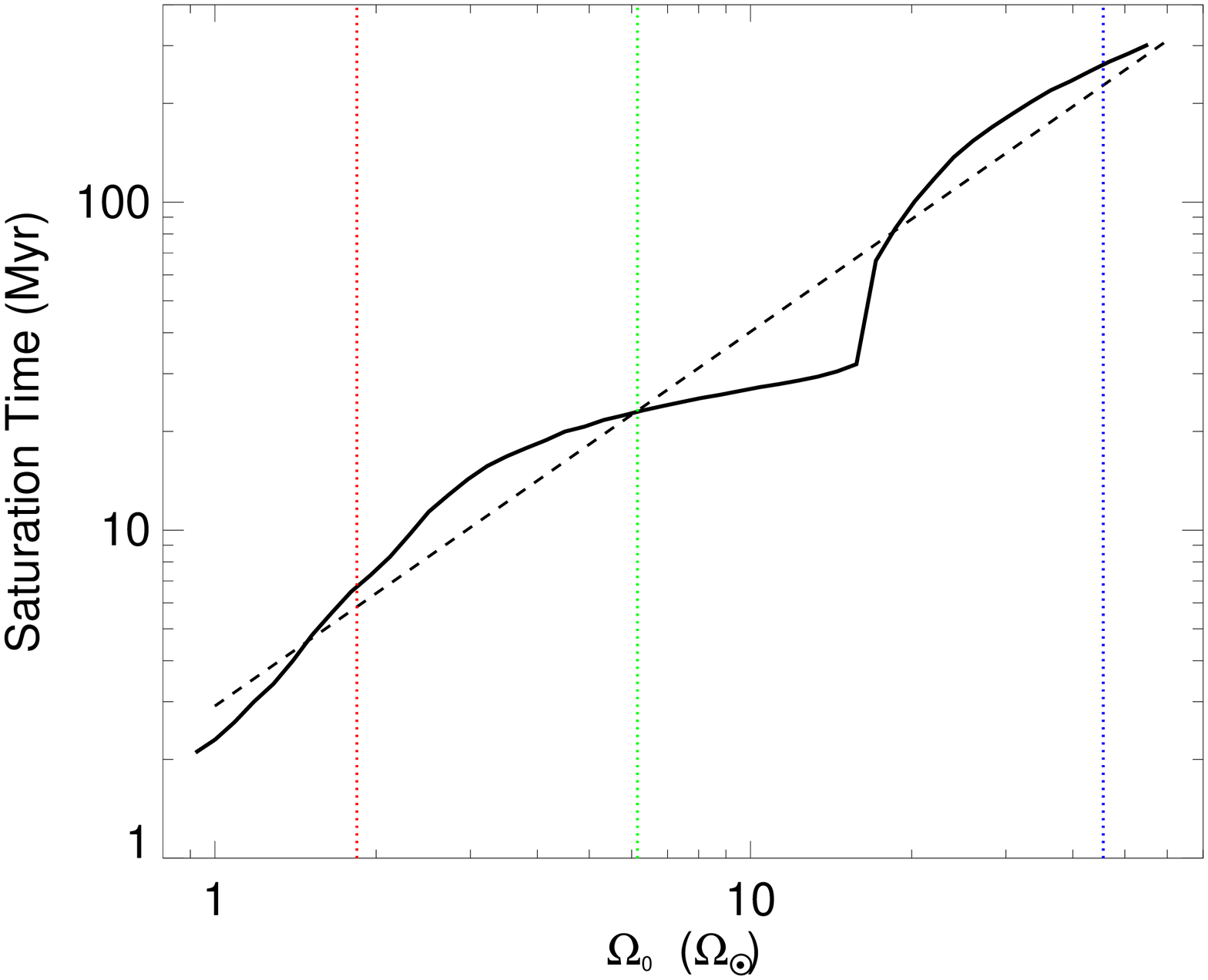}
\caption{Saturation time as a function of initial rotation rate $\Omega_0$.
To calculate each rotation track, we fit power laws to the core-envelope coupling timescale and disk-locking timescales for our 10th, 50th, and 90th percentile models of \mbox{$\tau_{\text{CE}}=38 \Omega_0^{-0.34}$} and \mbox{$t_\text{disk} = 13.5 \Omega_0^{-0.5}$}, where the timescales are given in Myr and $\Omega_0$ is in solar units.
The dashed line shows our best fit, given by Eqn.~\ref{tsat}, and the
vertical lines show the saturation times of the 10th, 50th, and 90th
percentile rotators.
}

\end{figure}

%
%
%
%
%
%
%
%
%
%
%
%

\section{Discussion}
The large differences in the evolutionary tracks, and therefore $L_{\rm X}$ and $L_{\rm EUV}$
values, make it necessary to reconsider critically the atmospheric erosion
of planets by high-energy radiation.
To first approximation, the thermal mass loss rate from a simple hydrogen
dominated planetary atmosphere, $\dot{M}_\text{pl}$, can be estimated
using the energy limited approach (\citealt{1981Icar...48..150W};
\citealt{2009A&A...506..399L}), where $\dot{M}_\text{pl}$ is proportional
to the incident stellar EUV flux for a given set of planetary parameters.
We therefore assume that $\dot{M}_\text{pl} \propto F_\text{EUV}$, where
$F_\text{EUV}$ is the EUV flux at the planetary orbit.
As an example, we consider the case of a 0.5~M$_\text{Earth}$ planet at
1~AU around a 1 M$_\odot$ star with an initial hydrogen atmosphere of $5
\times 10^{-3}$~M$_\text{Earth}$.
For this case, \citet{lammer14} calculated $\dot{M}_\text{pl}$ from the
atmosphere of $3.5 \times 10^{32} m_\text{H}$~s$^{-1}$ with $F_\text{EUV}
= 100$~erg~s$^{-1}$~cm$^{-2}$, where $m_\text{H}$ is the mass of a
hydrogen atom (see the fifth case in their Table~4).
We therefore assume that $\dot{M}_\text{pl} = 5.9 \times 10^6
F_\text{EUV}$, where $F_\text{EUV}$ is in erg~s$^{-1}$~cm$^{-2}$ and
$\dot{M}$ is in g~s$^{-1}$.

\begin{figure}
\includegraphics[angle=0,width=0.49\textwidth]{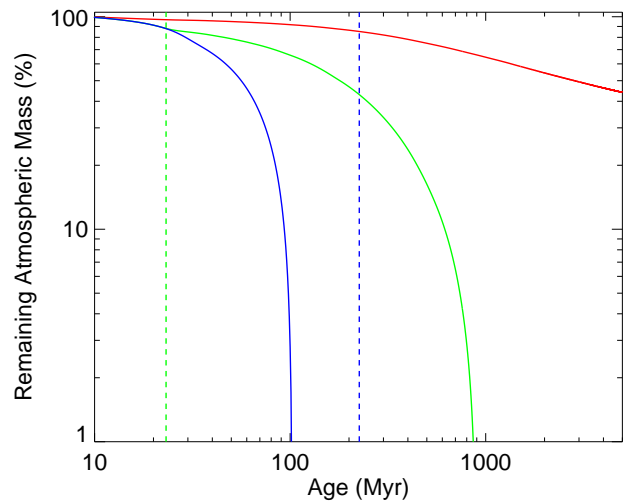}
\caption{
The evolution of the planetary atmospheric mass of a 0.5~M$_\text{Earth}$ planet orbiting a 1~M$_\odot$ star at 1~AU with an initial mass of \mbox{$5 \times 10^{-3}$~M$_\text{Earth}$}.  
The tracks correspond to planets orbiting stars that are on the 10th (\emph{red}), 50th (\emph{green}), and 90th (\emph{blue}) percentiles of the rotational distributions. 
The vertical lines show the stellar saturation times. 
}\label{fig:atmosphere}
\end{figure}

We show in Fig.~\ref{fig:atmosphere} the evolution of the planetary
atmospheric mass between 10~Myr and 5~Gyr assuming that the central star
follows the $L_\text{EUV}$ tracks shown in Fig.~\ref{fig:tracks}.
Although in all three cases, the planet at 10~Myr has identical
atmospheric masses, by 5~Gyr the atmospheric hydrogen contents are very
different.
Orbiting the slowly rotating star, the planet retains 45\% of its
initial atmosphere; orbiting the rapidly rotating star, the planet loses
its entire atmosphere within 100~Myr; orbiting the 50th percentile
rotator, the planet also loses its atmosphere, but this instead takes
almost a Gyr.
Although this is a simple calculation of a single example atmosphere, it
is sufficient to show that the star's initial rotation rate, and the
subsequent rotational evolution, is an important aspect that needs to be
properly considered when studying the evolution of terrestrial planetary
atmospheres.

%
%
%
%

\begin{acknowledgements}
The authors thank the referee, Nicholas Wright, for valuable comments.
LT was supported by an ``Emerging Fields'' grant of the
University of Vienna through the Faculty of Earth Sciences,
Geography and Astronomy.

CPJ, MG, and HL acknowledge the support of the FWF NFN project S11601-N16 
``Pathways to Habitability: From Disks to Active Stars, Planets and Life'', 
and the related FWF NFN subprojects S11604-N16 ``Radiation \& Wind Evolution 
from T Tauri Phase to ZAMS and Beyond'' and S11607-N16 ``Particle/Radiative Interactions with
Upper Atmospheres of Planetary Bodies Under Extreme Stellar
Conditions". This publication is supported 
by the Austrian Science Fund (FWF).
\end{acknowledgements}

\bibliographystyle{aa}
\bibliography{paperbib}


%
%
%
\end{document}